\documentclass[fleqn,usenatbib]{mnras}
\usepackage{newtxtext}
\usepackage{txfonts}

\usepackage[T1]{fontenc}
\usepackage{ae,aecompl}

\usepackage{graphicx}	

\title[Physical Inference from Fermi Blazars]{Physical Inference from the $\gamma$-ray, X-Ray and Optical Time Variability of a Large Sample of Fermi Blazars}

\author[A. Majumder et al.]
{Anwesh Majumder,$^{1}$\thanks{E-mail: anweshmajumder7@gmail.com}
Kaustav Mitra,$^{1,2}$
Ritaban Chatterjee,$^{1}$
C. M. Urry,$^{3,4}$
C. D. Bailyn$^{2}$ 
\newauthor and Prantik Nandi$^{1}$\thanks{Present address: S. N. Bose National Centre for Basic Sciences, JD Block, Sector III, Salt Lake City, Kolkata 700106, West Bengal, India.}
\\
\\
$^{1}$ Department of Physics, Presidency University, 86/1 College Street, Kolkata 700073, West Bengal, India \\
$^{2}$ Department of Astronomy, 52 Hillhouse Avenue, Steinbach Hall, Yale University, New Haven, CT 06511, USA\\ 
$^{3}$ Department of Physics, Yale University, 217 Prospect St, New Haven, CT 06511, USA \\ 
$^{4}$Yale Center for Astronomy and Astrophysics, Yale University, PO Box 208120, New Haven, CT 06520-8120, USA \\
}

\date{Submitted to MNRAS}

\pubyear{2018}

\begin{document}
\label{firstpage}
\pagerange{\pageref{firstpage}--\pageref{lastpage}}
\maketitle

\begin{abstract}
We present cross-correlation studies of $\gamma$-ray (0.1-300 GeV), X-ray (0.2-10 keV) and optical (R-band) variability of a sample of 26 blazars during 2008-2016. The light curves are from \textit{Fermi}-LAT, \textit{Swift-XRT}, and the Yale-SMARTS blazar monitoring program. We stack the discrete cross-correlation functions of the blazars such that the features that are consistently present in a large fraction of the sample become more prominent in the final result. We repeat the same analysis for two subgroups, namely, low synchrotron peaked (LSP) and high synchrotron peaked (HSP) blazars. We find that, on average, the variability at multiple bands is correlated, with a time lag consistent with zero in both subgroups. We describe this correlation with a leptonic model of non-thermal emission from blazar jets. By comparing the model results with those from the actual data we find that the inter-band cross-correlations are consistent with an emission region of size $\sim 0.1$ pc within the broad line region for LSP blazars. We rule out large changes of magnetic field ($> 0.5$ Gauss) across the emission region or small values of magnetic field ($\sim 0.2$ Gauss) for this population. We also find that the observed variability of the HSP blazars can be explained if the emission region is much larger than the distance to the broad line region from the central black hole. 
\end{abstract}




\section{Introduction}   \label{intro}
Blazars are a class of active galactic nuclei (AGN) that contain a bright relativistic jet pointed within a few degrees of our line of sight \citep{urr95}. Emission from the blazar jets is relativistically beamed and hence dominates over that from the other parts of the AGN, e.g., the accretion disk and emission line region. Variability in all wavebands, from radio to $\gamma$-rays, is a distinctive property of blazars. Their emitted flux fluctuates at all wavebands by a factor of a few at timescales of days to years, with the $\gamma$-ray and X-ray emission often varying dramatically within a few hours. The variations are red-noise-like, i.e., the amplitude is greater at longer than at shorter timescales. Variability at multiple wavebands can be used to investigate the location and mechanism of high-energy emission, and to probe the physics of launching, collimation and acceleration of jets.  

\begin{figure*}
	\includegraphics[trim={4cm 0 4cm 0}, height=11cm, width=16cm]{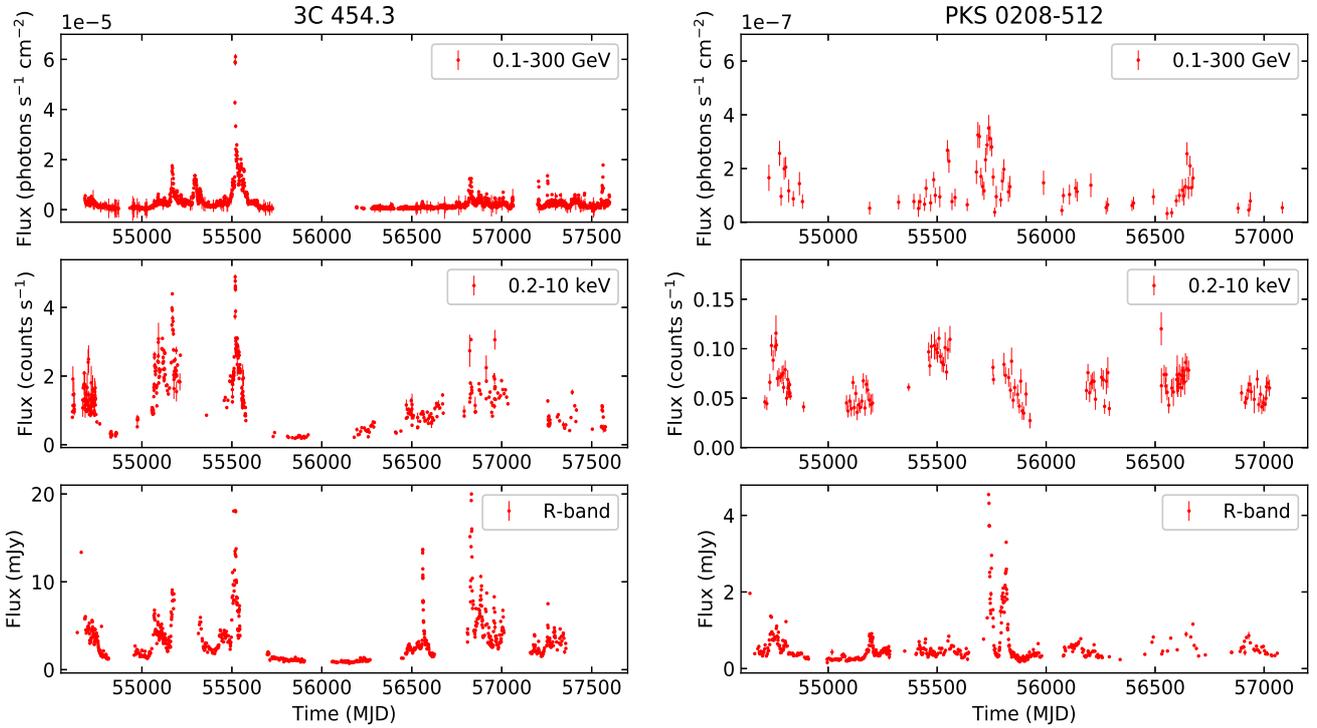}
	\caption{Optical (R$-$band), X-ray (0.2$-$10 keV) and $\gamma$-ray (0.1$-$300 GeV) light curves of FSRQs 3C 454.3 and PKS 0208-512.}
    \label{fig:light_curve}
\end{figure*}

Highly relativistic electrons in the magnetic field of the jet produce the radio-optical emission via synchrotron radiation  \citep{bre81,urr82,imp88,mar98}.  In the so called ``Leptonic Model'' for jet emission, the same distribution of electrons is responsible for up-scattering lower energy photons to X-ray and $\gamma$-ray energies through the inverse-Compton (IC) process. The lower energy ``seed photons'' may be the synchrotron photons from the jet itself, in which case it is called synchrotron self-Compton \citep[SSC;][]{mar92,chi02,arb05} or external to the jet, e.g., from the accretion disk, broad line region or the dusty torus, which is thus termed external Compton (EC) process \citep{sik94,cop99,bla00,der09}. 
Optical synchrotron emission is produced by electrons of Lorentz factor $\gamma \sim 10^3$, assuming a magnetic field ($B$) of a few Gauss. Emission at 1 GeV may be produced by electrons at similar energies by the up-scattering of infrared (IR) photons from the torus. X-rays, on the other hand, may be produced by very high-energy ($\gamma \sim 10^5$) electrons through the synchrotron process or by electrons of lower energy ($\gamma \sim 10^2$) through the SSC or EC processes. As a result, optical and GeV emission may be generated by the highest energy electrons, and hence may exhibit the fastest variability. X-rays, on the other hand, if generated by electrons at lower energies, should vary more slowly. Emission at optical, X-ray and GeV energies defines the shape of the spectral energy distribution (SED), and hence unambiguous knowledge about the variability at these bands and their inter-relation can provide stringent constraints on SED models. Therefore, it is imperative to study the cross-correlations among the optical, X-ray, and $\gamma$-ray wave bands in a large sample of blazars to compare with the predictions of the Leptonic Model.

While numerous cross-correlation studies have been carried out by many authors \citep[e.g.,][]{cha08,jor10,bot10,bon12,hay12}, those mostly concentrated on individual objects with well-sampled light curves. Now a large sample of blazars is available because the Large Area Telescope (LAT) onboard the \textit{Fermi} Gamma-Ray Space Telescope has been observing the $\gamma$-ray sky since its launch in 2008. Supporting multi-wavelength campaigns have followed \textit{Fermi}-detected blazars at a range of wavebands across the electromagnetic spectrum. Therefore, it is now possible to investigate the nature of the multi-band cross-correlations for a large sample of blazars. For example, \citet{fuh14} stacked cross-correlation functions (CCF) of the GeV and radio variations of a large sample of blazars and found that the time delays between the fluctuations at those two wave bands vary monotonically with the wavelength of the radio emission used. In order to obtain this result, instead of looking at the CCF of individual blazars, they stacked the CCFs of all blazars of the sample such that the features consistently present among the blazars become prominent in the stacked' CCF. Results from such analyses provide important constraints on the spectral and temporal behavior of blazars that viable theoretical models have to reproduce. For this reason, we here collect optical, X-ray and $\gamma$-ray light curves of $26$ blazars to compute their average cross-correlation. We then construct a theoretical model that generates light curves at different energy bands. By comparing the CCFs of observed light curves with those from our theoretical model, we are able to constrain the typical physical parameters of blazar jets.

In {\S}\ref{data} we describe the optical, X-ray, and $\gamma$-ray data that we use in this work while in {\S}\ref{cc_analysis} we present the cross-correlation analysis. We describe our model and theoretical predictions in {\S}\ref{model}. In {\S}\ref{discussion}, we compare observational and theoretical results and discuss their implications.  The results are summarized in {\S}\ref{summary}. 

\section{Data}    \label{data}

\begin{figure*}
	\includegraphics[trim={2.5cm 0 2.5cm 0}, height=9cm, width=18cm]{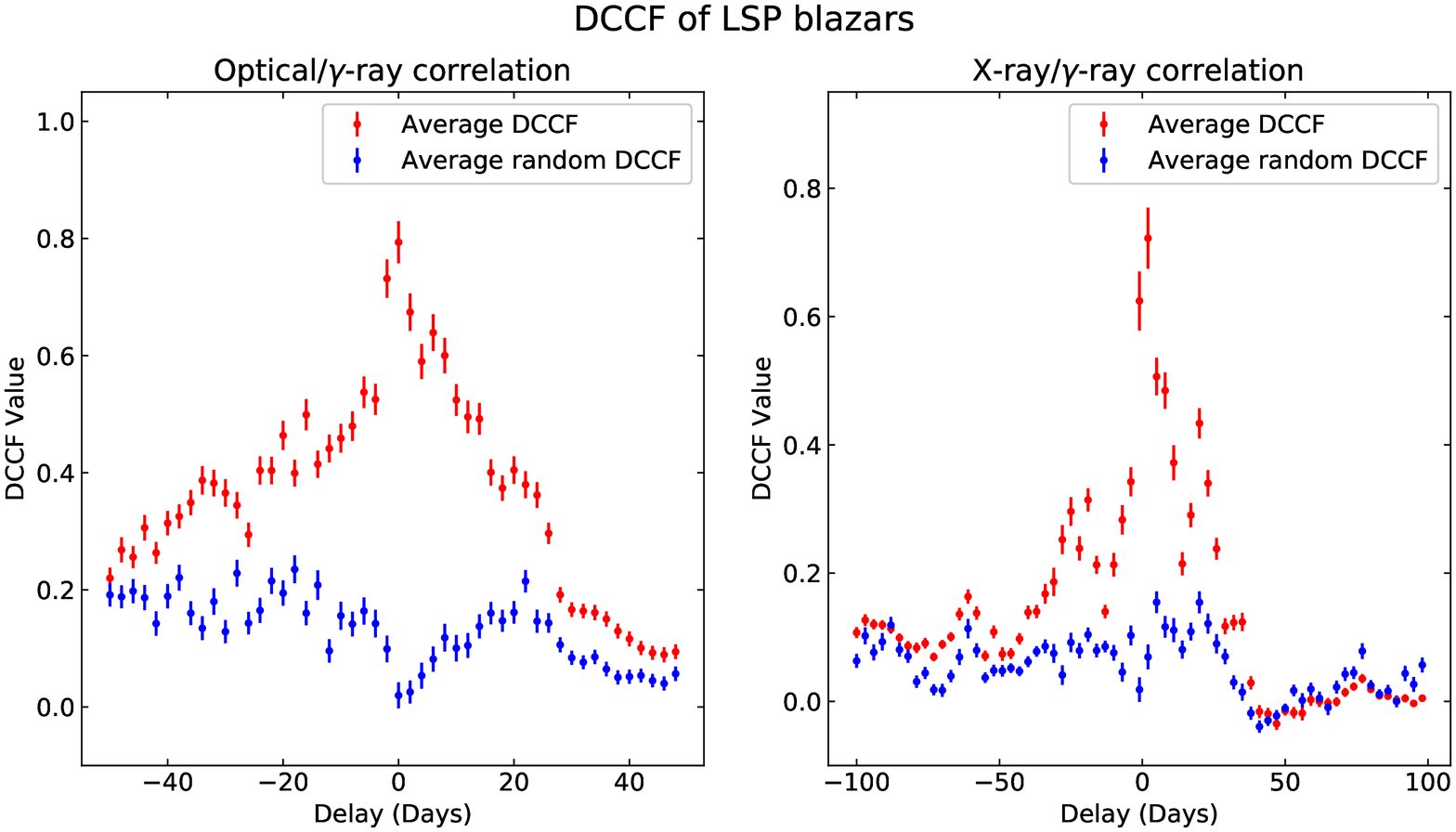}
	\caption{Average discrete cross-correlation functions (DCCFs) of well-monitored FSRQs and LBLs. {\it Left:} Stacked optical/$\gamma$-ray DCCF for 3C 273, 3C 279, 3C 454.3, PKS 0208-512, PKS 0235+164, PKS 0402-362, PKS 0426-380, PKS 0454-234, PKS 0537-441, PKS 1244-255, PKS 1424-41, PKS 1510-089, PKS 2142-75, PKS 2326-502, CRATESJ 0531-4827 and OJ 287. {\it Right:} Stacked X-ray/$\gamma$-ray DCCF for 3C 279, 3C 454.3, CTA 102, PKS 0208-512, PKS 0235+164, PKS 0716+714, PKS 1222+216, PKS1424-41, PKS 1510-089, PKS 1633+382 and BL Lacertae. The time delay is defined as positive if the variations at the higher frequency lead those at the lower frequency.}
    \label{fig:lsp_cc}
\end{figure*}

We retrieve the GeV data from the FSSC website\footnote{https://fermi.gsfc.nasa.gov/ssc/}. More specifically we use the all-sky weekly data files as provided by the \textit{Fermi} team to facilitate the data analyses of a large number of blazars, as is required here\footnote{https://fermi.gsfc.nasa.gov/ssc/data/analysis/scitools/LAT\_weekly\_allsky.html}. We select the data using the energy range: $0.1-300$ GeV, time range: the year 2008 to 2018, and the coordinates of the sources (see Tables \ref{tab:targets1} and \ref{tab:targets2} for sources). We carry out unbinned likelihood analysis of the data using the most recent version of the \textit{Fermi} data analysis tool, namely, {\tt Fermitools} and with the {\tt P8R2\_SOURCE\_V6} instrument response function (IRF). We select events classified as {\tt evclass}~=~128 and {\tt evtype}~=~3 within the region of interest (ROI) of 10$^{\circ}$ around the co-ordinates of the blazar for further analysis. We use the filter ``DATA\_QUAL$==$1" and ``LAT\_CONFIG$==$1" to select good time intervals (GTI). We include nearby sources as obtained from the 4-year LAT catalog \citep[3FGL;][]{3FGL} given in \textit{gll\_psc\_v16.fit} to create a model of all possible $\gamma$-ray sources that may contribute to the flux observed. We model the spectra of the blazars with a power-law and keep all parameter values fixed at their respective catalog value except the source of interest for which we keep both the index and the normalization free while for other sources within $3^{\circ}$ of the target the latter is kept free. We include the Galactic and extragalactic diffuse emission and istropic background emission in the model using the templates \textit{gll\_iem\_v06} and \textit{iso\_P8R2\_SOURCE\_V6\_v06}, respectively. We consider a source to be significantly detected in a given time bin if the corresponding value of the Test Statistic (TS) $>25$. We generate the light curves by carrying out the above analyses for each time bin of interest throughout the entire interval. We use 1-day time bins for all the blazars in our sample except PKS 0537-441, 0208-512, OJ 287, 0235+164, PKS 1244-255, for which we use 7-day bins because the number of significant detections with 1-day binning is low. 

We use the X-ray light curves from the \textit{Swift-XRT} monitoring program of \textit{Fermi}-LAT sources of interest \citep{str13}. As part of this observing program, usually blazars that are bright in the $\gamma$-rays are monitored regularly by \textit{Swift-XRT}, often with daily pointings of $\sim1$ ks, while sources that are not flaring in the $\gamma$-rays at the time are observed occasionally. Therefore, the X-ray light curves are irregularly sampled and absence of X-ray data does not necessarily imply non-detection. 

We obtain the optical R-band light curves from the Yale-SMARTS blazar monitoring program. The observations were carried out by the ANDICAM instrument on the SMARTS 1.3m telescope located at CTIO, Chile, and thus covers declinations south of $\sim 10^{\circ}$. As in the X-ray, sources were followed when they were flaring in $\gamma$-rays and/or bright or flaring in the optical/IR, so coverage is not complete. For details of data acquisition, calibration and data reduction procedures, see \citet{bon12}. 

Our sample consists of the blazars that have been observed at least 50 times in each of the above wavebands over the period 2008-2016. A total of 26 blazars have this level of coverage in either $\gamma$-ray and optical or $\gamma$-ray and X-ray frequencies. In particular, we have $16$ blazars for optical/$\gamma$-ray and $16$ blazars for X-ray/$\gamma$-ray correlation. While many blazars in our sample are similar to 3C 454.3 (Figure \ref{fig:light_curve}, left panel), which was observed very frequently in all three bands, with high significance, there are a few blazars like PKS 0208-512 (Figure \ref{fig:light_curve}, right panel), which have been observed sporadically in one or two bands and have larger uncertainties.

\section{Cross correlation analysis}   \label{cc_analysis}

We use the discrete cross-correlation function \citep[DCCF;][]{ede88} to study the 
inter-band variability, e.g., optical/$\gamma$-ray and X-ray/$\gamma$-ray light curves, of the blazars in our sample. While calculating the DCCF, we employ a threshold of a minimum of 10 data points in each bin of time-delay in order to ensure that the DCCF values are statistically significant. We calculate the optical/$\gamma$-ray DCCF with time-delay bins of $2$ days because the $\gamma$-ray light curves of most of the blazars in our list are well sampled. For X-ray/$\gamma$-ray DCCF we use a bin size of $3$ days as the X-ray light curves from \textit{Swift-XRT} are less well sampled with uncertainties of DCCF too large for shorter time delay bins.

\begin{table}
	\centering
	\caption{List of blazars used in optical/$\gamma$-ray DCCF analysis and their time delay values. Column 1 lists each source and column 2 is the classification according to its spectral energy distribution. Column 3 lists the optical/$\gamma$-ray delay with uncertainties in days. The time delay is defined as positive if the variations at the higher frequency wave band lead those at the lower frequency.}
	\label{tab:targets1}
	\begin{tabular}{ccc} 
		\hline
		Sources	&  	Type	&	Delay	\\ 
		\hline
		3C 273 				&		LSP		&	-5 $\pm$ 1		\\
		3C 279 				&		LSP		&	3 $\pm$ 2		\\
		3C 454.3 			& 		LSP		&	-4.3 $\pm$ 0.6	 \\
		PKS 0208-512		&		LSP		&	2 $\pm$ 2  \\
	    PKS 0235+164 		&		LSP		&	16 $\pm$ 1 \\
	    PKS 0402-362 		&		LSP		&	1.8 $\pm$ 0.5 \\
	    PKS 0426-380 		&		LSP		&	5.9 $\pm$ 0.9 \\
	    PKS 0454-234 		&		LSP		&	-4.8 $\pm$ 0.7 \\
	    PKS 0537-441 		& 		LSP		&	84 $\pm$ 2 \\
	    PKS 1244-255    	&		LSP		&	-0.8 $\pm$ 0.8 \\
	    PKS 1424-41 		&		LSP		&	0 $\pm$ 1 \\
	    PKS 1510-089		&		LSP    	&   No correlation \\
	    PKS 2142-75			&		LSP		&	4 $\pm$ 1  	\\
	    PKS 2326-502		&		LSP		&	-2 $\pm$ 1  \\
	    CRATESJ 0531-4827	&		LSP 	& 	34 $\pm$ 2 \\
	    OJ 287 				&		LSP		& 	No correlation \\
		\hline
	\end{tabular}
\end{table}

The cross-correlation function in one blazar may be dominated by a single feature in a long-term light curve. Furthermore, due to the transient nature of blazars, cross-correlation results for one source are often not representative of the entire population. Therefore, we follow \citet{fuh14} in averaging the cross-correlation function of multiple blazars such that the features that are consistently present in a large fraction of the sample become more prominent in the final result. Due to various observational constraints in blazar monitoring as discussed in {\S}\ref{data}, light curves at these three bands are not regularly sampled. The peak value of the correlation function of irregularly sampled data points over a limited time range may be artificially low or high. In order to ensure that blazars which have high peak correlation value do not dominate the final averaged DCCF, we normalize the DCCF of each blazar by dividing by its peak value before performing the average weighted by their respective uncertainties. 

\begin{table}
	\centering
	\caption{List of blazars used in X-ray/$\gamma$-ray DCCF analysis and their time delay values. Column 1 lists each source and column 2 is the classification according to its spectral energy distribution. Column 3 lists the X-ray/$\gamma$-ray delay with uncertainties in days. The time delay is defined as positive if the variations at the higher frequency wave band lead those at the lower frequency.}
	\label{tab:targets2}
	\begin{tabular}{ccc} 
		\hline
		Sources	& 	Type	&	Delay \\
		\hline
		3C 279 			&	LSP		&	2 $\pm$ 1 \\
		3C 454.3 		&   LSP		&	-0.3 $\pm$ 0.7 \\
		CTA 102         &   LSP		&	No correlation \\
		PKS 0208-512    &   LSP		&	No correlation \\
		PKS 0235+164    &   LSP		&	No correlation \\
		PKS 0716+714    &   LSP		&	16 $\pm$ 2 \\
		PKS 1222+216    &   LSP		&	No correlation \\
		PKS 1424-41     &   LSP		&	-6 $\pm$ 2 \\
		PKS 1510-089    &   LSP		&	No correlation \\
		PKS 1633+382    &   LSP		&	2.4 $\pm$ 0.8 \\
		BL Lacertae     &   LSP		&	No correlation \\
		1ES1959+650		&	HSP		&	No correlation \\
		3C 66A		&	HSP		&	No correlation \\
		PKS 2155-304	&	HSP		&	4 $\pm$ 2 \\
		Mrk 421		&	HSP		&	No correlation \\
		Mrk 501		&	HSP		&	No correlation \\
		\hline
	\end{tabular}
\end{table}

\begin{figure}
	\includegraphics[trim={2.7cm 0 2.7cm 0}, height=6cm, width=\columnwidth]{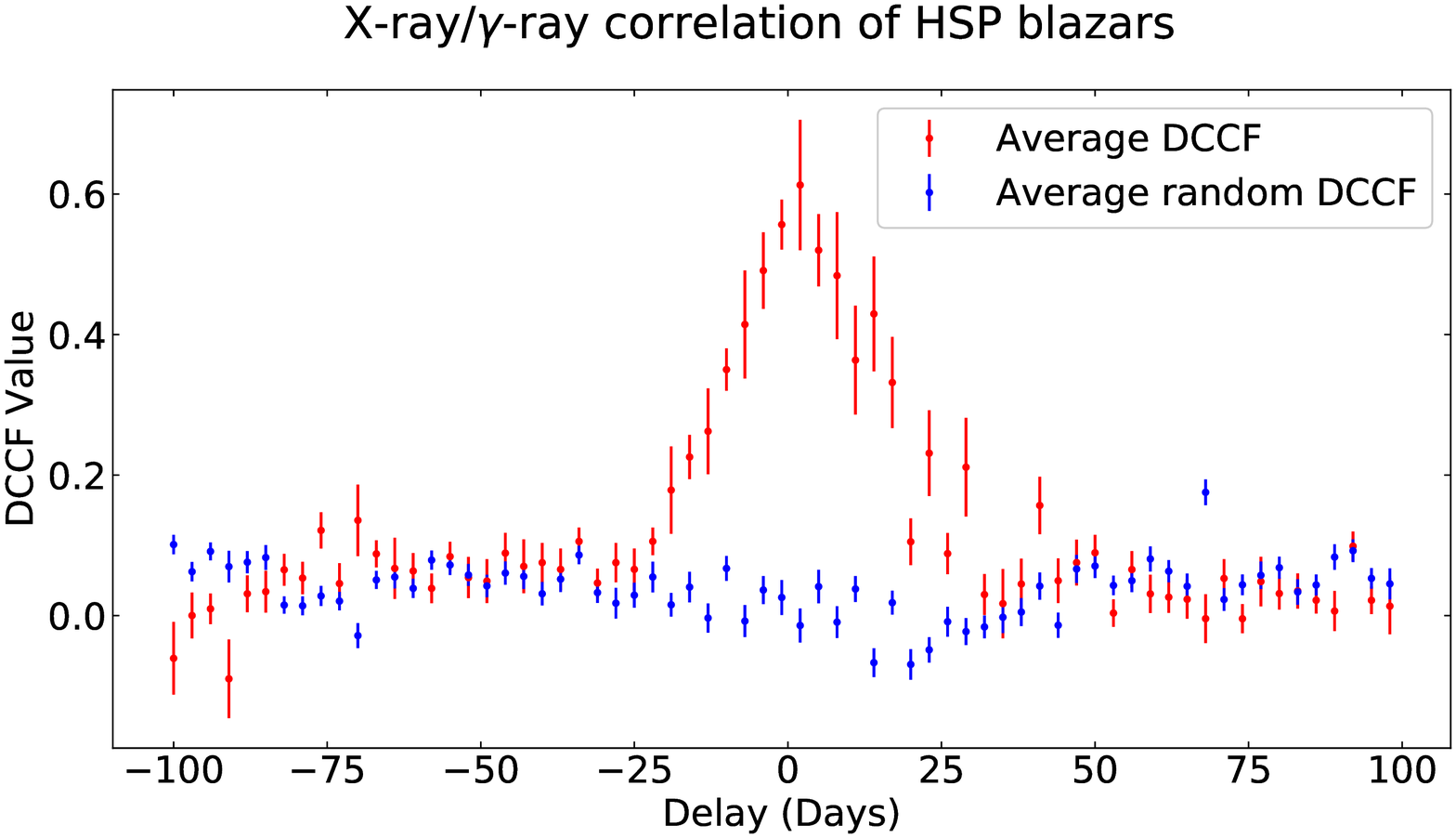}
	\caption{Average discrete cross-correlation function (DCCF) of the X-ray/$\gamma$-ray light curves of HBLs in the sample (3C 66A, 1ES1959+650, PKS 2155-304, Mrk 421, Mrk 501). Positive time delays mean variations at higher frequency lead those at lower frequency. 
	}
    \label{fig:hsp_cc}
\end{figure}

We divide our sample into low-frequency synchrotron peaked (LSP) blazars and high-frequency synchrotron peaked (HSP) blazars. The LSP blazars consists of flat spectrum radio quasars (FSRQ) and low-frequency peaked BL Lac (LBL) objects while HSP blazars consist of high-frequency peaked BL Lac (HBL) objects. All $16$ optical light curves and $11$ X-ray light curves in our sample are of LSP blazars while $5$ X-ray light curves are of HSP blazars. Tables \ref{tab:targets1} and \ref{tab:targets2} lists all the blazars used in this study and their corresponding DCCF time delays with errors. The time delay and its uncertainty are found by finding a best-fit Gaussian profile to the DCCFs. For the optical/$\gamma$-ray correlation, we find that 8 LSP blazars show a positive delay, 1 has zero delay, 5 have negative delays and 2 show no significant correlation. In our sample, all but 3 LSP blazars have delay within $\pm$ $6$ days; the exceptions are CRATESJ 0531-4827, PKS 0235+164, and PKS 0537-441, which have larger time delays ($> 10$ days). The $\gamma$-ray data of CRATESJ 0531-4827 are irregularly sampled and hence this may contribute to the large time delay. We find that PKS 0537-441 exhibits a delay of more than 50 days. Such DCCF peaks at extremely large delays may occur when one compares emission from different regions in the jet at different times. For the X-ray/$\gamma$-ray correlation, we find 3 blazars with positive delays, 2 blazars with negative delays, and 6 blazars showing no significant correlation. Next, we stack the individual DCCFs of LSP and HSP blazars separately. In Figure \ref{fig:lsp_cc}, we present the average optical/$\gamma$-ray (left panel) and X-ray/$\gamma$-ray (right panel) DCCF of LSP blazars in our sample. In both the cases the average DCCF has a strong peak near zero time delay. In addition, we cross-correlate the $\gamma$-ray light curve of each blazar with the optical and X-ray light curve of the other blazars and determine the average of the above DCCF values in each time-delay bin. We plot these random DCCF along with the stacked DCCF in order to check whether the stacked correlation result is significant. It can be seen that the stacked DCCF clearly lies above the random DCCF values in most time-delay bins. A positive value of delay implies that variations at the higher frequency band lead those at the lower frequency. We find the delay for the peak of stacked correlation to be $-1 \pm 2$ days. For the X-ray/$\gamma$-ray correlation, we find that the delay peak is located at $-1 \pm 3$ days.

\begin{figure}
	\includegraphics[trim={2.7cm 0 2.7cm 0}, height=6cm, width=\columnwidth]{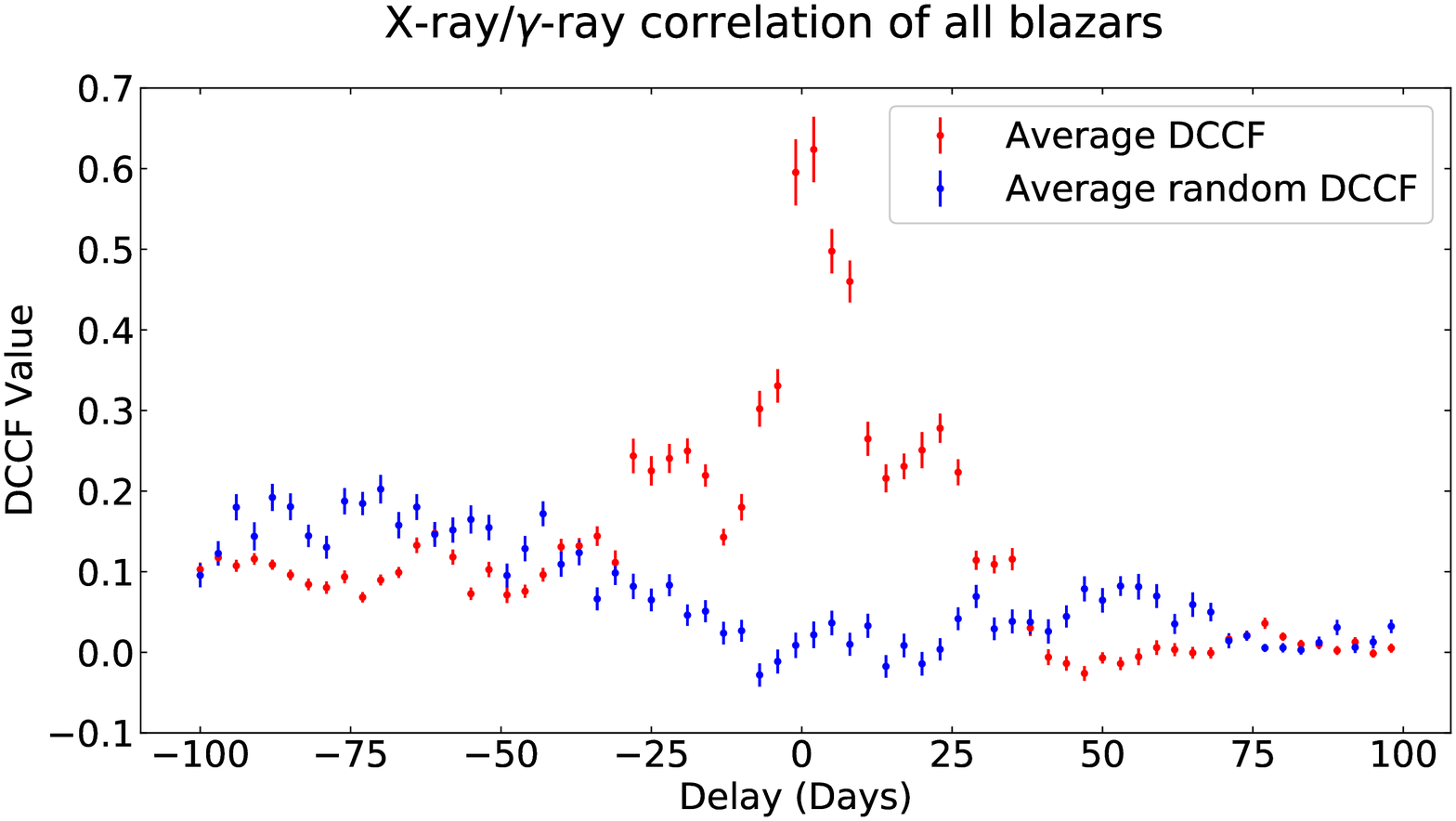}
	\caption{Average discrete cross-correlation function (DCCF) of the X-ray/$\gamma$-ray light curves of all blazars in the sample. 
	Positive time delays mean variations at higher frequency lead those at lower frequency.
	}
    \label{fig:total_cc}
\end{figure}  

In Figure \ref{fig:hsp_cc}, we present the X-ray/$\gamma$-ray correlation of HSP blazars in our sample. In this case, there is only $1$ blazar with positive delay (PKS 2155-304) while other $4$ blazars show no significant correlation. We find the peak delay to be $2 \pm 2$ days. Finally, the X-ray/$\gamma$-ray DCCF of all the blazars, both LSP and HSP blazars, are stacked (Figure \ref{fig:total_cc}) to study the collective behavior of these populations. We find that the DCCF peak is at a delay of $-1$ $\pm$ $2$ days between X-ray/$\gamma$-ray.

\section{Theoretical modeling}    \label{model}
We model the non-thermal flares in blazars using a computer code that assumes an emission region of rectangular cross-section. The region is located $\sim 0.1$ pc from the central black hole. It is divided into $50$ cells from which radiation reaches the observer. The cross-sectional area of the emission zone is chosen such that it subtends an angle of few degrees at a distance of $\sim 0.1$ pc from the central black hole. It is assumed that the electron evolution in one cell does not affect the evolution in another cell. The strength of the magnetic field may vary from cell to cell according to $B\sim r^{-1}$, where $r$ is the distance of the cell from the base of the jet. A shock front passes through the emission region at a speed of $v_{sh}$, which accelerates the electrons. In our model this is achieved by instantly injecting each cell with electrons having a power law energy distribution $N(\gamma) = N_0\gamma^{-s}$ within a range $\gamma_{\rm min}$ to $\gamma_{\rm max}$, where $\gamma$ is the energy in the units of the electron rest mass. These high energy electrons then lose their energy by synchrotron radiation and inverse-Compton (IC) scattering while the shock moves to the next cell. The seed photons that are up-scattered in the IC process includes both synchrotron photons as well as photons that are external to the jet (e.g., broad line region (BLR) photons and radiation from torus).

\begin{table}
	\centering
	\caption{Input parameters held constant to model the non-thermal emission.}
	\label{tab:table1}
	\begin{tabular}{lc} 
		\hline
		Parameters & Values\\
		\hline
		Minimum electron Lorentz factor, $\gamma_{min}$ & $50$\\
		Speed of shock, $v_{sh}$ & $c/\sqrt{3}$ \\
		Size of the emission region, (pc)		&	$0.1$	\\
		Normalization constant of electron distribution, $N_0$	&	$10^{55}$		\\
		Energy of BLR photons, $h\nu_{BLR}$ (eV)		&	$10$\\
		Frequency of torus radiation, $\nu_{torus}$ (Hz)   &   $10^{13}$ \\ 
		Initial slope of electron distribution, $s$ & $2.5$ \\ 
		Luminosity of accretion disk, $L_{D}$ (erg s$^{-1}$) & $10^{45}$ \\
		\hline
	\end{tabular}
\end{table}

We compute the radiation profile without taking into consideration the time delay of seed photons (light travel time) to scattering with electrons or the Klein-Nishina effect. While calculating the SSC emission in a given cell, we add the contribution of seed photons from all other cells weighted by the inverse-squared of their distance. We take into account the Doppler boosting of the radiation from each cell to the observer's frame when calculating the light curves at different band. This is given by:
\begin{equation}
	\frac{\partial L_{\nu}}{\partial \Omega} = D^3\frac{\partial L'_{\nu'}}{\partial \Omega'},
	\label{luminosity}
\end{equation}
where the Doppler factor ($D$) due to the motion of the plasma in the jet is given by: 
\begin{equation}
	D = \frac{1}{\Gamma_{jet}(1-\beta_{jet} \cos \theta)}.
\end{equation} 
Here, $\theta$ is the angle between the cells and the line of sight to the observer. For small cross-section (as in our case), this angle can be approximately set to zero. Equation \ref{luminosity} relates the monochromatic power per unit solid angle between the observer's frame (unprimed quantities) and the jet frame (primed quantities). In addition, the time dilation effect due to the motion of the jet plasma is taken into account. Radiation from all zones is then summed to calculate the final light curves as a function of time in the observer's frame. For calculating the nature of the synchrotron and inverse-Compton radiation (from both synchrotron and external seed photons) and the evolution of the electron energy distribution due to radiation cooling, we employ the scheme of \citet{mod03}. The external photon density $u_{ext}'$ is due to emission from the BLR and torus. The BLR and torus radiation density in the jet frame as a function of distance from the black hole is given by \citep{hay12}: 
\begin{equation}
	u_{BLR}'(r) = \frac{\epsilon_{BLR}\Gamma_{jet}^2L_D}{3\pi r_{BLR}^2c[1+(r/r_{BLR})^{\beta_{BLR}}]},
	\label{blr}
\end{equation}
\begin{equation}
	u_{torus}'(r) = \frac{\epsilon_{torus}\Gamma_{jet}^2L_D}{3\pi r_{torus}^2c[1+(r/r_{torus})^{\beta_{torus}}]},
	\label{torus}
\end{equation}
\noindent
where $\epsilon_{BLR} = 0.1$ and $\epsilon_{torus} = 0.01$ represent fraction of accretion disk luminosity reprocessed into emission lines and into hot dust radiation, respectively, $L_D$ is the accretion disk luminosity, and $c$ is the speed of light. We set $\beta_{BLR} = 3$ and $\beta_{torus} = 4$ \citep{hay12}. The distance to BLR from central black hole, $r_{BLR}$, is taken to be $\sim$ $0.9$ pc. The distance to the torus from the central black hole, $r_{torus}$, on the other hand, is taken to be $\sim$ $10$ pc. In Equations (\ref{blr}) and (\ref{torus}), $r$ is the distance to the cell from the central engine.

\begin{figure}
	\includegraphics[trim={2.7cm 0 2.7cm 0}, height=6cm, width=\columnwidth]{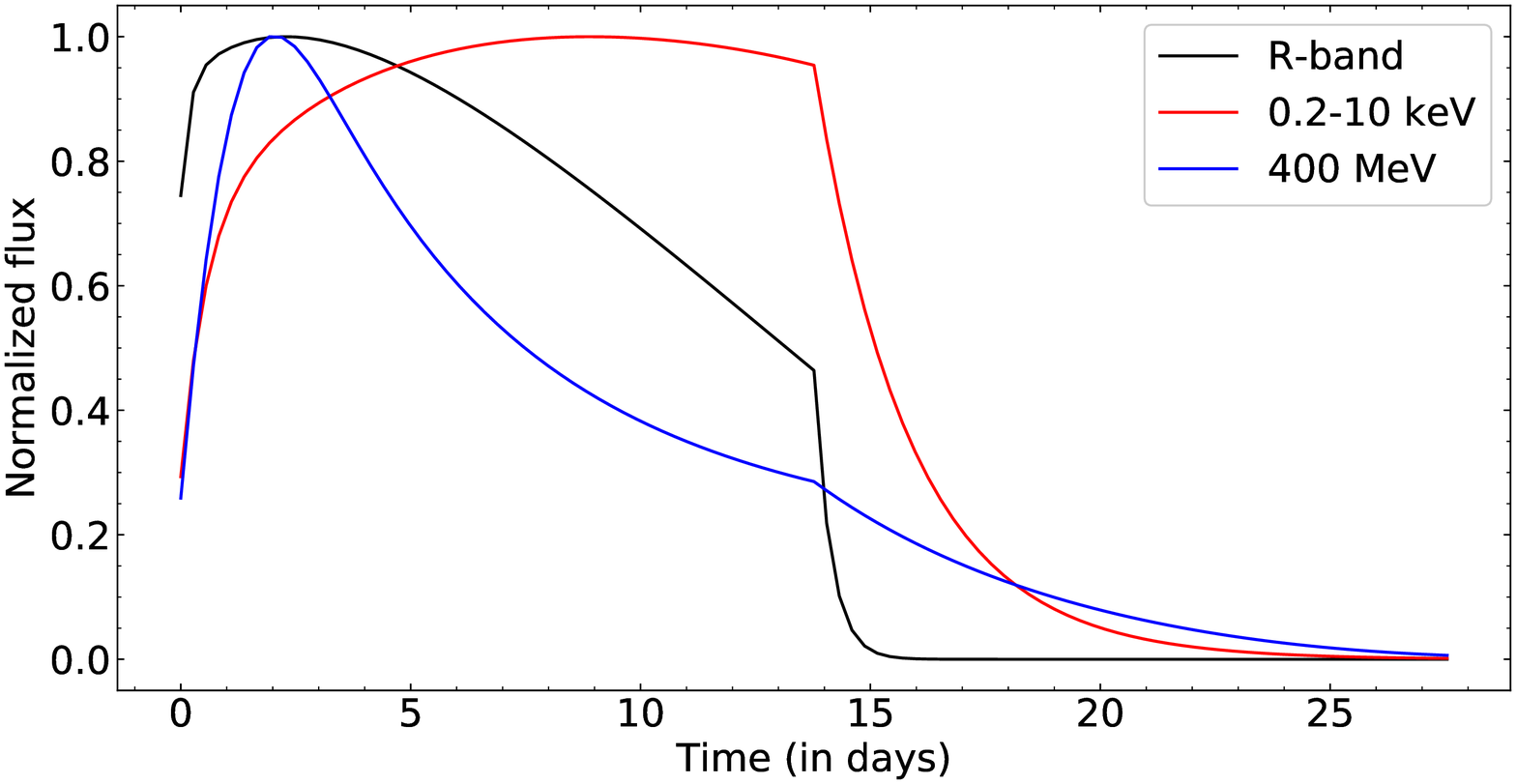}
	\caption{Multi-wavelength variability simulated using our model. We set $L_D = 1\times 10^{45}$ erg s$^{-1}$ and the size of the emission region $\sim 0.2$ pc. In this case, the light curves are much wider than $10$ days. For these two example runs, we set $\Gamma_{jet} = 15$, $\gamma_{max} = 3000$, $B_0 = 0.6$ G and $B_f = 0.3$ G.}
    \label{fig:example1}
\end{figure} 

\begin{table*}
	\centering
	\caption{Input parameters varied to model non-thermal emission of FSRQs.}
	\label{tab:table2}
	\begin{tabular}{lccc} 
		\hline
		Parameters & Initial Value & Final Value & Step size\\
		\hline
		Maximum electron Lorentz factor, $\gamma_{max}$ & $1000$ & $4000$ & $1000$\\
		Bulk Lorentz factor of the jet plasma, $\Gamma_{jet}$ & $10$ & $20$ & $5$\\
		Magnetic field at one end of the emission region, $B_0$ (G)	& $0.2$ & $1.0$	& $0.2$\\ 
		Magnetic field at the other end of the emission region, $B_f$ (G) & $0.1$ & $B_0-0.1$ & $0.2$	\\ 
		\hline
	\end{tabular}
\end{table*} 

Our model requires input parameters such as the bulk Lorentz factor of the jet ($\Gamma_{jet}$), magnetic field at both ends of the emission region ($B_0$ and $B_f$), energy of photons external to the jet, etc. In order to interpret the results discussed in {\S}\ref{cc_analysis}, we generate the light curves by varying some of those parameters and observe which set of parameters 
agrees with the cross-correlation results obtained before. The parameters in our model that are held constant are described in Table \ref{tab:table1}. We assume the BLR and the torus emit at the UV and infrared wavelengths, respectively. 

\begin{figure*}
	\includegraphics[trim={2.5cm 0 2.5cm 0}, height=10cm, width=18cm]{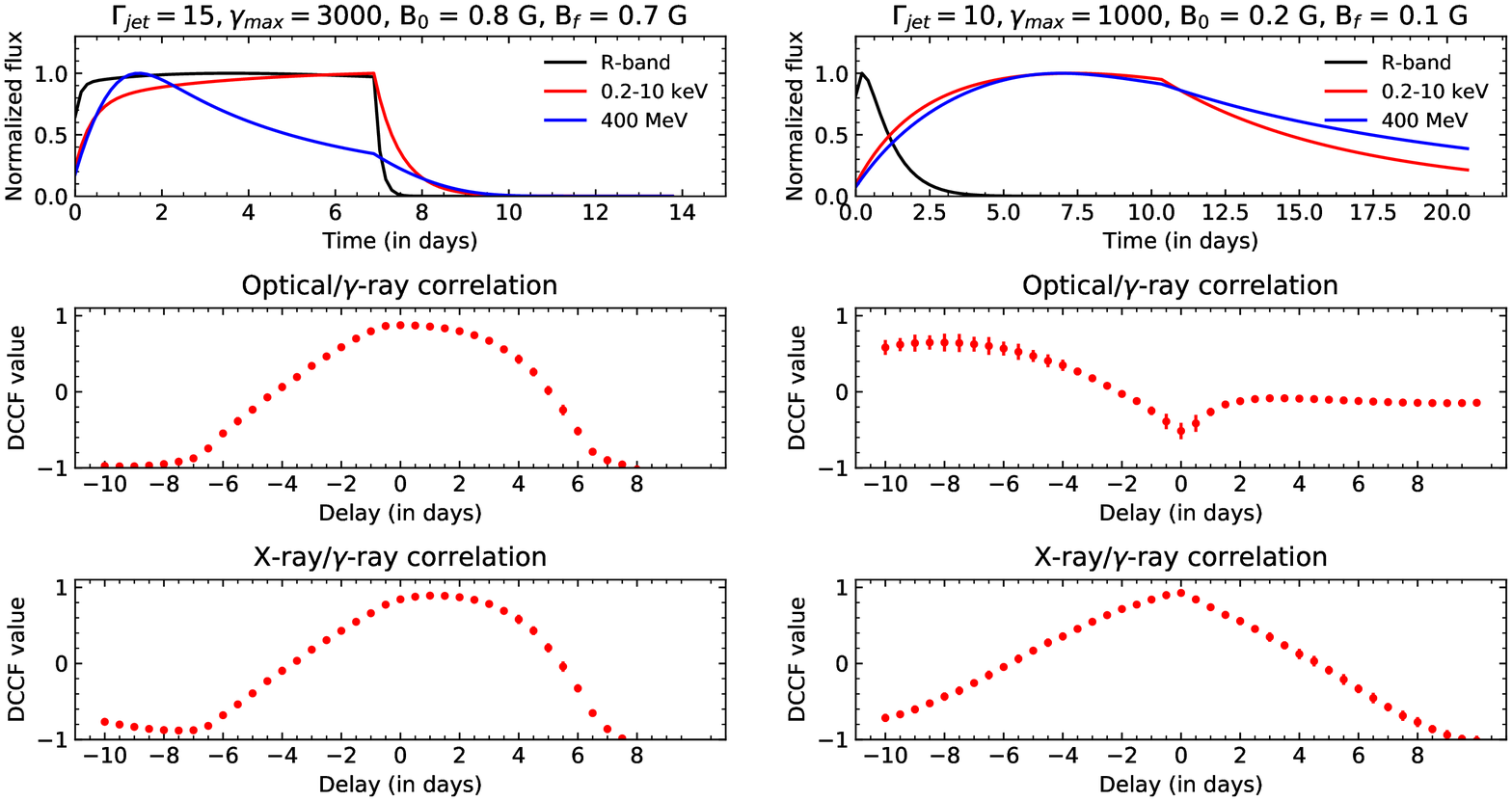}
	\caption{Model light curves for two different parameter sets. The normalized flux reaches unity when the shock has energized all the cells. The time delay is defined as positive if variations at the higher frequency wave band lag those at the lower frequency. The optical/$\gamma$-ray and X-ray/$\gamma$-ray DCCF peaks for $\Gamma_{jet} = 15$, $\gamma_{max} = 3000$, $B_0 = 0.8$ G and $B_f = 0.7$ G are consistent with the results described in {\S}\ref{cc_analysis}. For $\Gamma_{jet} = 10$, $\gamma_{max} = 1000$, $B_0 = 0.2$ G and $B_f = 0.1$ G, optical/$\gamma$-ray and X-ray/$\gamma$-ray DCCF peaks are inconsistent with the results described in {\S}\ref{cc_analysis}.}
    \label{fig:model}
\end{figure*}

\begin{figure*}
	\includegraphics[trim={3cm 0 3cm 0}, height=10cm, width=16.5cm]{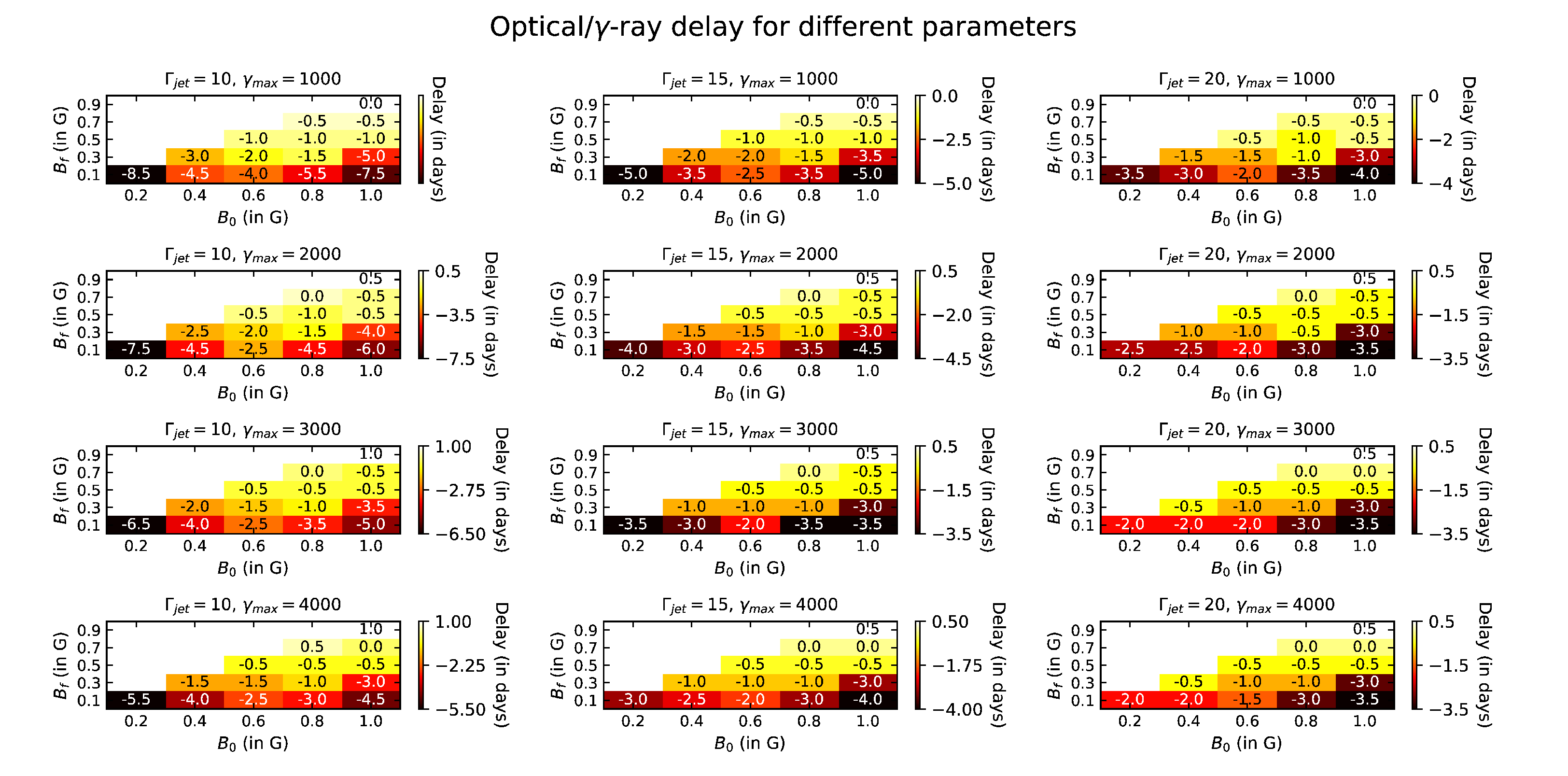}
	\caption{Optical/$\gamma$-ray delay (in days) for different parameters in the parameter space. A negative time delay corresponds to variations at the higher frequency wave band lagging those at the lower frequency.}
    \label{fig:op_gamma_result}
\end{figure*}
 
In order to generate the light curves so that we can compare with the results of the LSP blazars in {\S}\ref{cc_analysis}, we set $L_D = 10^{45}$ erg s$^{-1}$ which is the typical accretion disk luminosity for FSRQ blazars \citep[e.g., see BLR luminosity in Table 4 of][ which is $\sim10$\% of disk luminosity]{ghi11}. Next, we set the emission region size appropriately so that, given the parameter space we consider, the timescale of the flares is not longer than $10$ days, i.e., $\sim 0.1$ pc. This is consistent with the time delays we find in {\S}\ref{cc_analysis}, which are a few days. If we instead set the size of the emission region to $\sim 0.2$ pc, we get light curves similar to that of Figure \ref{fig:example1}, where the width of the flares is more than $10$ days.

We vary the maximum Lorentz factor of electrons ($\gamma_{max}$), the bulk Lorentz factor of the jet plasma ($\Gamma_{jet}$) and the magnetic field at both ends of the emission region ($B_0$ and $B_f$), as described in Table \ref{tab:table2}. We restrict the value of $\gamma_{max}$ in the parameter space to $4000$ to avoid any significant contribution from the Klein-Nishina regime. 
Even for $\gamma_{max} = 4000$ and $\Gamma_{jet} = 20$, the highest possible photon energy in the electron rest frame is $\Gamma_{jet}\gamma_{max} h\nu_{ext} = 800$ keV, comparable to the rest mass of the electron. In addition, we find that the $R$-band and 400 MeV light curves become flat-topped for high values of $\gamma_{max}$. This is due to the fact that when $\gamma_{max}$ is high, optical and $400$ MeV emission are produced by a comparatively lower energy population of electrons. Due to the radiation cooling effects the number density of lower energy electrons that can produce R$-$band and 400 MeV emission is fairly stable, which causes the long timescales for the flares. Hence, the emission in this band continues to increase, saturates after some time, and remains the same till the shock has left the emission region. It is difficult to identify the peak of the DCCF with an uncertainty no larger than a few days between a pair of such flat-topped flares, which are more than $5$ days wide. However, DCCF with sharp peak is present in the observational result. Hence, we restrict the exploration of the parameter space for which R$-$band and $400$ MeV radiation are emitted by relatively higher energy population of electrons. 

\begin{figure*}
	\includegraphics[trim={3cm 0 3cm 0}, height=10.5cm, width=16.5cm]{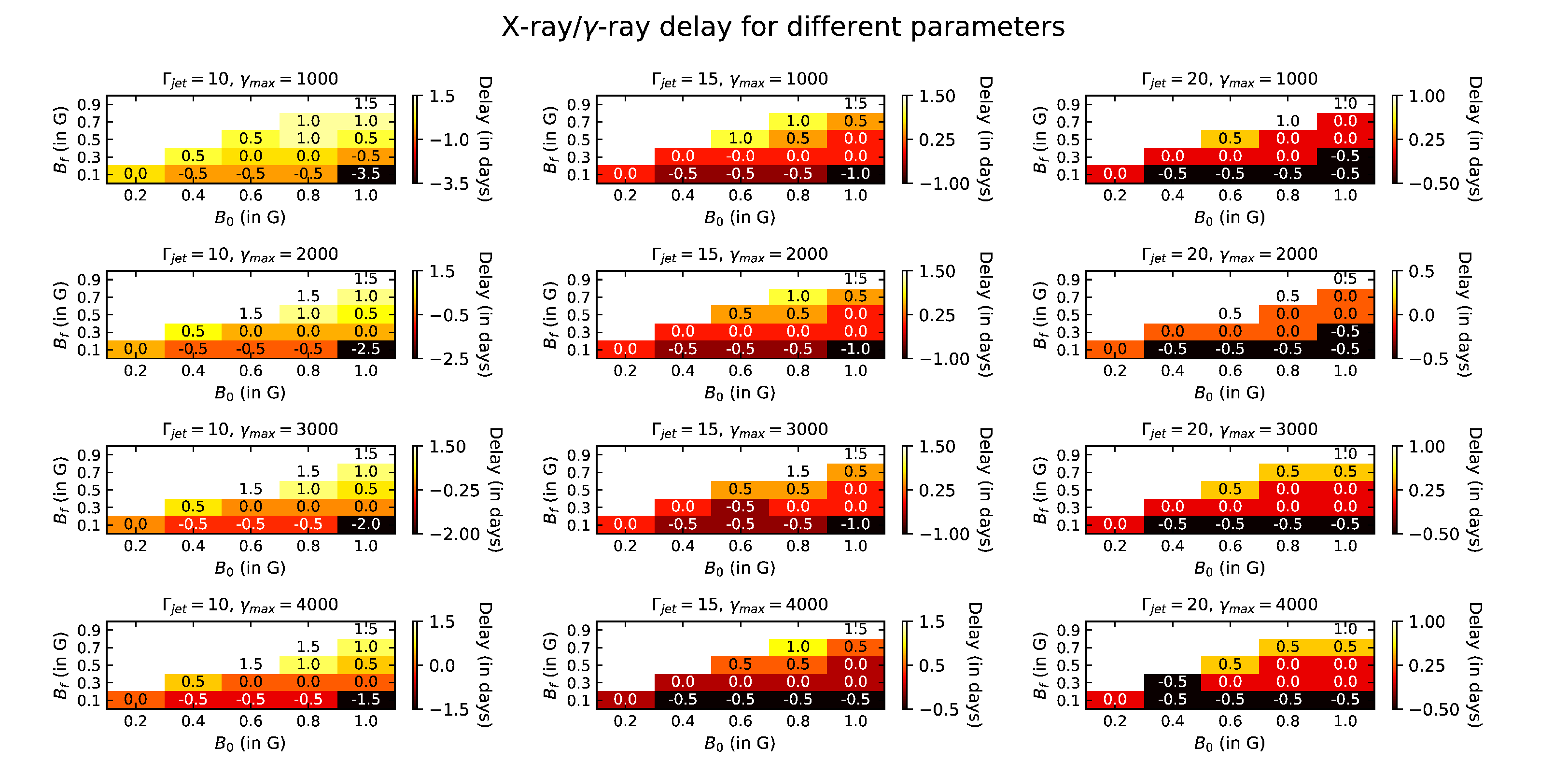}
	\caption{X-ray/$\gamma$-ray delay (in days) for different parameters in the parameter space. A negative time delay corresponds to variations at the higher frequency wave band lagging those at the lower frequency.}
   \label{fig:x_gamma_result}
\end{figure*} 

\begin{figure*}
	\includegraphics[trim={2.5cm 0 2.5cm 0cm}, height=10cm, width=18cm]{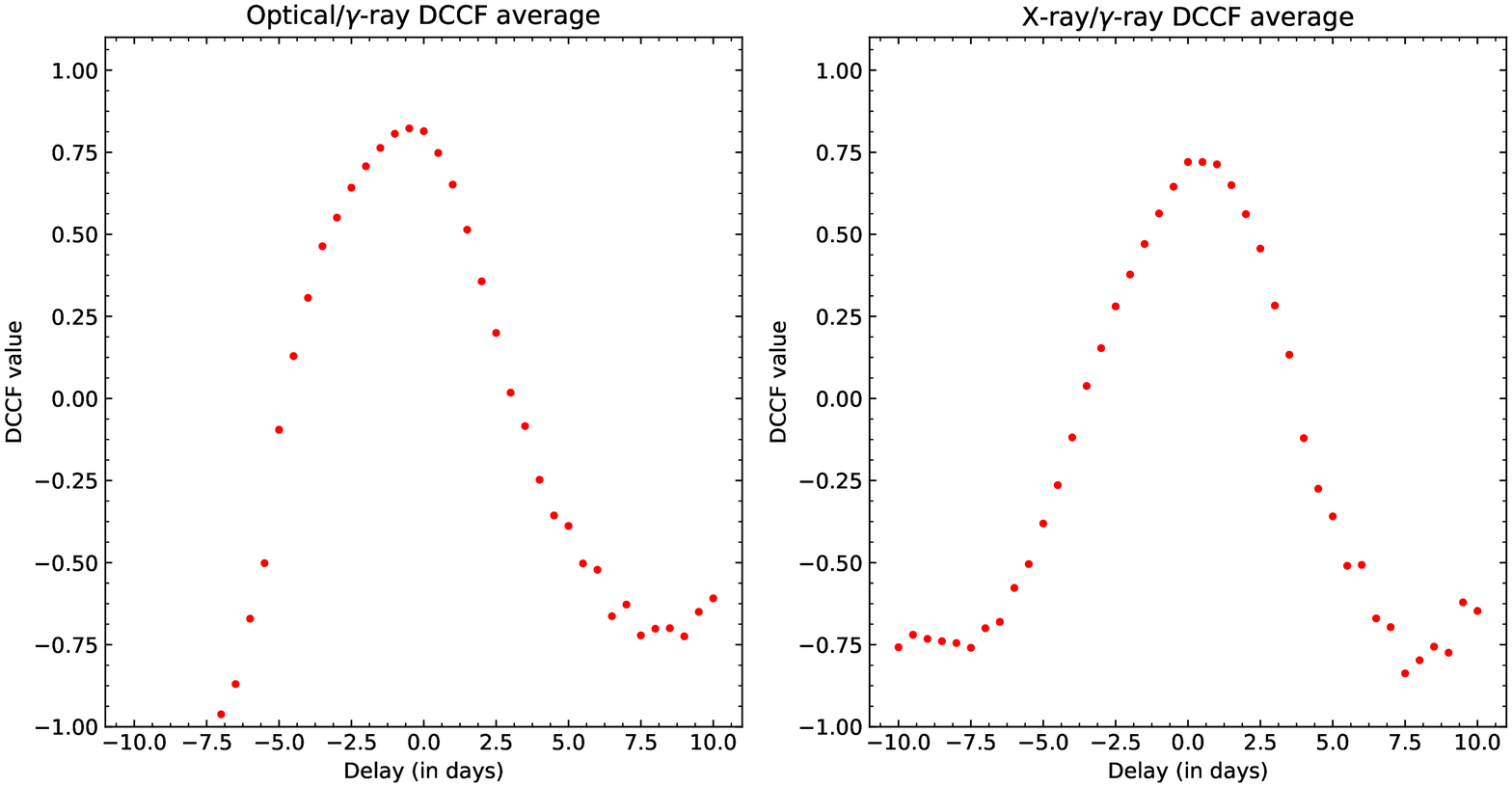}
	\caption{Average cross-correlation results for parameters in which optical/$\gamma$-ray and X-ray/$\gamma$-ray delay as shown in Figure \ref{fig:op_gamma_result} and \ref{fig:x_gamma_result} are consistent with results obtained in {\S}3.}
    \label{fig:model_avg}
\end{figure*}

With these parameters, we run our model to create different light curves at $R-$band, $0.2-10$ keV and $400$ MeV energies. Two typical results from our model and cross-correlation analysis for optical/$\gamma$-ray and X-ray/$\gamma$-ray variability are shown in Figure \ref{fig:model}. In the first case, we set $\Gamma_{jet} = 15$, $\gamma_{max} = 3000$, $B_0 = 0.8$ G, $B_f = 0.7$ G. For the second case, we set $\Gamma_{jet} = 10$, $\gamma_{max} = 1000$, B$_0 = 0.2$ G, B$_f = 0.1$ G. The rest of the parameters are as specified in Table \ref{tab:table1}.

In general, the emission at the R$-$band, $0.2-10$ keV and $400$ MeV energies increases as more and more cells are energized by the shock. The optical synchrotron radiation starts to decrease before the shock leaves the emission region as the magnetic field falls off inversely with distance. As a result, radiation from the latter emission zones contributes less to the total synchrotron radiation. Furthermore, as the rate of loss of energy increases rapidly with electron energy, \begin{large}$d\gamma/dt'$\end{large} $\propto \gamma^2$, the high energy electrons that produce optical radiation quickly lose their energy. Thus, the number of electrons that produce optical radiation is reduced. These two effects combine to produce an optical light curve that starts to decrease before the shock has left the emission zone. $\gamma$-ray emission similarly decreases as the external photon density decreases with distance and is produced by relatively high energy electrons. However, the population of electrons that produces optical emission has higher energy compared to those producing 400 MeV emission. As a result, electrons producing $R-$band emission lose their energy earlier than those producing $\gamma$-rays. Hence, optical emission reaches its maximum and decays earlier compared to the $\gamma$-ray emission. The SSC radiation at $0.2-10$ keV energy, on the other hand, is produced due to the scattering of synchrotron photons of different energies by electrons which have different Lorentz factors. Thus, although high energy electrons lose their energy quickly, $0.2-10$ keV radiation can still be produced by the scattering of higher energy synchrotron photons by lower energy electrons. Due to this, the SSC radiation continues to increase as more and more electrons are energized by the shock. When the shock leaves the emission region, the SSC radiation decreases as the electrons lose their remaining energy. Overall, our simple model predicts that R-band and $\gamma$-ray light curves should have sharper flares while the X-ray flares are broader. 

Furthermore, we cross correlate the optical/$\gamma$-ray and X-ray/$\gamma$-ray light curves simulated above. In Figure \ref{fig:model} (left), it is clear that the optical/$\gamma$-ray cross-correlation peak is within $-1$ $\pm$ $2$ days and X-ray/$\gamma$-ray cross-correlation peak is within $-1$ $\pm$ $3$ days. Hence, the parameters used to simulate these LSP blazar light curves are favored by the results discussed in {\S}\ref{cc_analysis}. On the other hand, in Figure \ref{fig:model} (right), the optical/$\gamma$-ray cross-correlation peak is not within $-1$ $\pm$ $2$ days and hence the parameters used to simulate these light curves are disfavored by the results in {\S}\ref{cc_analysis}. We present the time delay in the optical/$\gamma$-ray DCCF for the entire range of values in our parameter space in Figure \ref{fig:op_gamma_result} and that for the X-ray/$\gamma$-ray DCCF in Figure \ref{fig:x_gamma_result}.

Although different blazar jets have different values of parameters $B_0$, $B_f$, $\Gamma_{jet}$ and hence different time delays between $\gamma$-ray/X-ray/optical light curves, we are interested in the general behaviour of the blazars. Hence, we stack the correlation results of pairs of light curves simulated with only those sets of parameters for which the delay shown in Figure \ref{fig:op_gamma_result} and \ref{fig:x_gamma_result} is consistent with observational values obtained in {\S}\ref{cc_analysis}. The stacking is done by calculating the average of DCCF values weighted by their uncertainties in each time bin. The result is shown in Figure \ref{fig:model_avg}. We find that the optical/$\gamma$-ray stacked DCCF show a delay of $-0.9 \pm 1.0$ days while that for the X-ray/$\gamma$-ray stacked DCCF is $0.5 \pm 0.7$ days. 

\begin{figure*}
	\includegraphics[trim={2.5cm 0 2.5cm 0cm}, height=10.5cm, width=18cm]{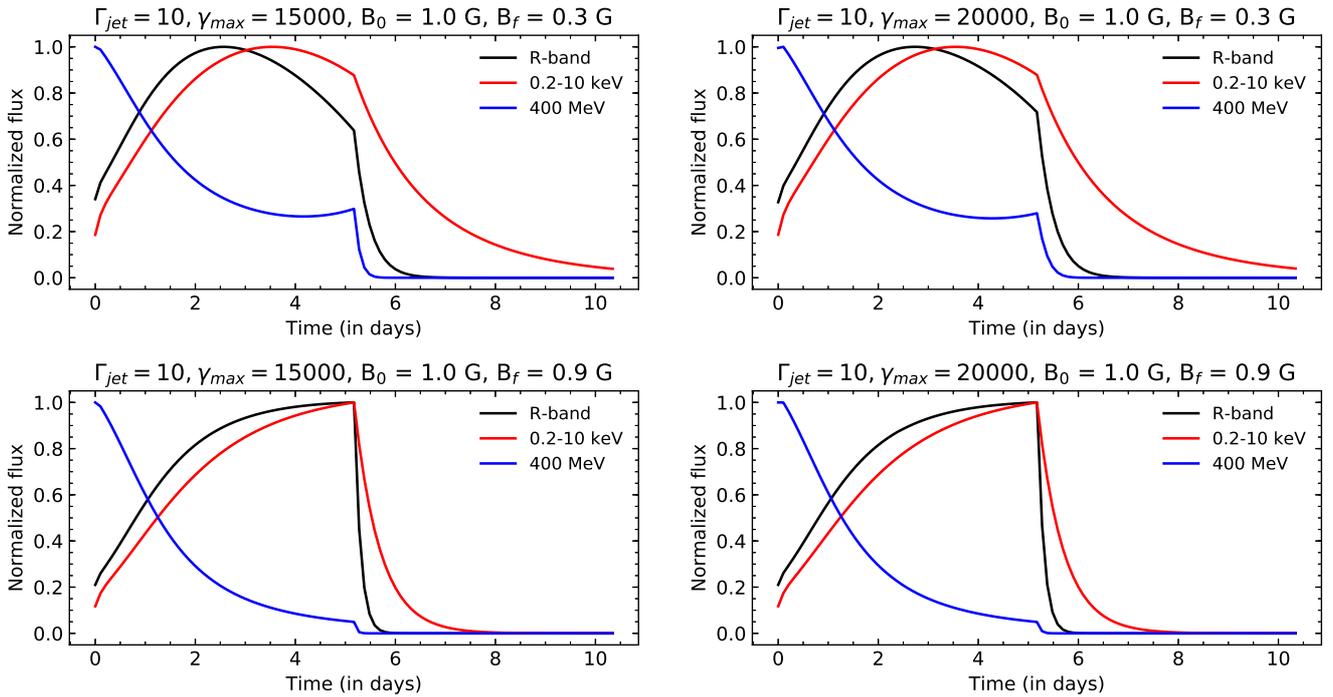}
	\caption{Light curves of HSP blazars simulated by our model. The normalized flux reaches unity when the shock has energized all the cells. The BLR luminosity is chosen to be $10^{42}$ erg s$^{-1}$, distance of the BLR from the central black hole to be $0.01$ pc and the size of the emission region to be $0.1$ pc.}
    \label{fig:hsp_model}
\end{figure*}

In order to simulate light curves that match the general shape of the observed variability of HSP blazars, we set the luminosity of the BLR to be $L_{BLR} = \epsilon_{BLR}L_D = 10^{42}$ erg s$^{-1}$ (see Equation \ref{blr}). We set the distance to the BLR from the central engine to $\sim 0.01$ pc. The delay between the observed optical/$\gamma$-ray and X-ray/$\gamma$-ray variations is a few days, which implies that the size of the emission region is a few times $0.1$ pc. Therefore, a significant part of the emission region will be beyond the BLR. When $r >> r_{BLR}$, from Equation \ref{blr}, it is expected that the energy density falls off rapidly. We show the results from four possible combination of parameters that can produce interesting results in Figure \ref{fig:hsp_model}. The plot shown in the upper left panel is the result of using the parameters $\Gamma_{jet} = 10$, $\gamma_{max} = 15000$, $B_0 = 1.0$ G and $B_f = 0.3$ G in the simulation. This demonstrates the scenario when there is a sharp variation of the magnetic field. In the lower left panel, on the other hand, the parameters used are $\Gamma_{jet} = 10$, $\gamma_{max} = 15000$, $B_0 = 1.0$ G and $B_f = 0.9$ G. This demonstrates the scenario for a small variation of the magnetic field. We repeat the same exercise with $\gamma_{max} = 20000$ in the right panels. For all of these cases, we set the size of the emission region to be $0.1$ pc extending from $0.01$ pc (i.e., from the BLR) to $0.11$ pc. Near one of the ends of the emission region, which is closer to the central engine, the external energy density is dominated by the BLR photons while it is dominated by the torus photons near the other end.

\begin{table*}
	\centering
	\caption{Optical/$\gamma$-ray and X-ray/$\gamma$-ray delays for different input parameters of the model to simulate BL Lacs. The size of the emission region is chosen to be $0.1$ pc and $\Gamma_{jet}$ to be $10$. A positive time delay corresponds to variations at the higher frequency wave band leading those at the lower frequency. The luminosity of the BLR, $L_{BLR}$, is taken to be $10^{42}$ erg s$^{-1}$.}
	\label{tab:table3}
	\begin{tabular}{ccccc} 
		\hline
		$\gamma_{max}$ & $B_0$ & $B_f$ & Optical/$\gamma$-ray delay (days) & X-ray/$\gamma$-ray delay (days)\\
		\hline
		$15000$     &     $1.0$ & $0.3$ & $1.5$			& 			  		$2.5$\\
		$15000$     &     $1.0$ & $0.9$ & $3.0$			& 			  		$3.5$\\
		$20000$    &     $1.0$ & $0.3$ & $2.0$			& 			  		$2.5$\\
		$20000$    &     $1.0$ & $0.9$ & $3.0$			& 			  		$3.5$\\
		\hline
	\end{tabular}
\end{table*}

In the top two panels of Figure \ref{fig:hsp_model}, the $\gamma$-ray emission decreases rapidly as the energy density of the external radiation falls off quickly with distance for $r > r_{BLR}$ (see Equation \ref{blr}). The optical emission is generated by low energy electrons and hence their cooling rate is small compared to the high energy electrons. As a result, the number of electrons that can produce optical emission increases as higher energy electrons lose their energy and as more cells are energized by the shock. Towards the downstream end of the emission region, the value of the magnetic field is quite low compared to $1.0$ G and hence the emission from these regions do not contribute much to the overall optical flux. Thus, the optical emission starts to decrease before the shock leaves the emission region. In the bottom two panels, on the other hand, the value of the magnetic field at the emission region is still comparable to $1.0$ G. As a result, the synchrotron radiation from these regions still contributes significantly to the overall optical flux and hence the emission continues to increase till the shock leaves the emission region. The SSC radiation in both of these cases continues to increase for reasons discussed previously. These are the primary effects that determine the nature of the light curves for the chosen scenario. In this case, the light curves are not very sensitive to the value of $\gamma_{max}$. The detailed values of the optical/$\gamma$-ray and X-ray/$\gamma$-ray delay are given in Table \ref{tab:table3}. For the chosen parameters, the X-ray/$\gamma$-ray delay value is consistent with the observed value of $2 \pm 2$ days.  

\section{Discussion}    \label{discussion}

In this section, we compare the theoretical results from our model with the observational results discussed in {\S}\ref{cc_analysis}. From the delays shown in Figures \ref{fig:op_gamma_result} and \ref{fig:x_gamma_result} for LSP blazars, we find that large changes in the magnetic field across the emission region ($> 0.5$ G) and very small values of the magnetic field (e.g., $0.2$ G) are consistently ruled out by our model. Therefore, we conclude that when the emission region is relatively close to the BLR and and it has a size $\sim0.1$ pc, blazar jets should neither have sharp changes in the value of magnetic field ($> 0.5$ G) across a distance of $\sim 0.1$ pc, nor should they have small magnetic field values such as $\sim 0.2$ G at such distances. The observed nature of variability for HSP blazars, on the other hand, can be explained if there is significant \textbf{jet} emission beyond the BLR.

\begin{figure}
	\includegraphics[trim={2.7cm 0 2.7cm 0}, height=6cm, width=\columnwidth]{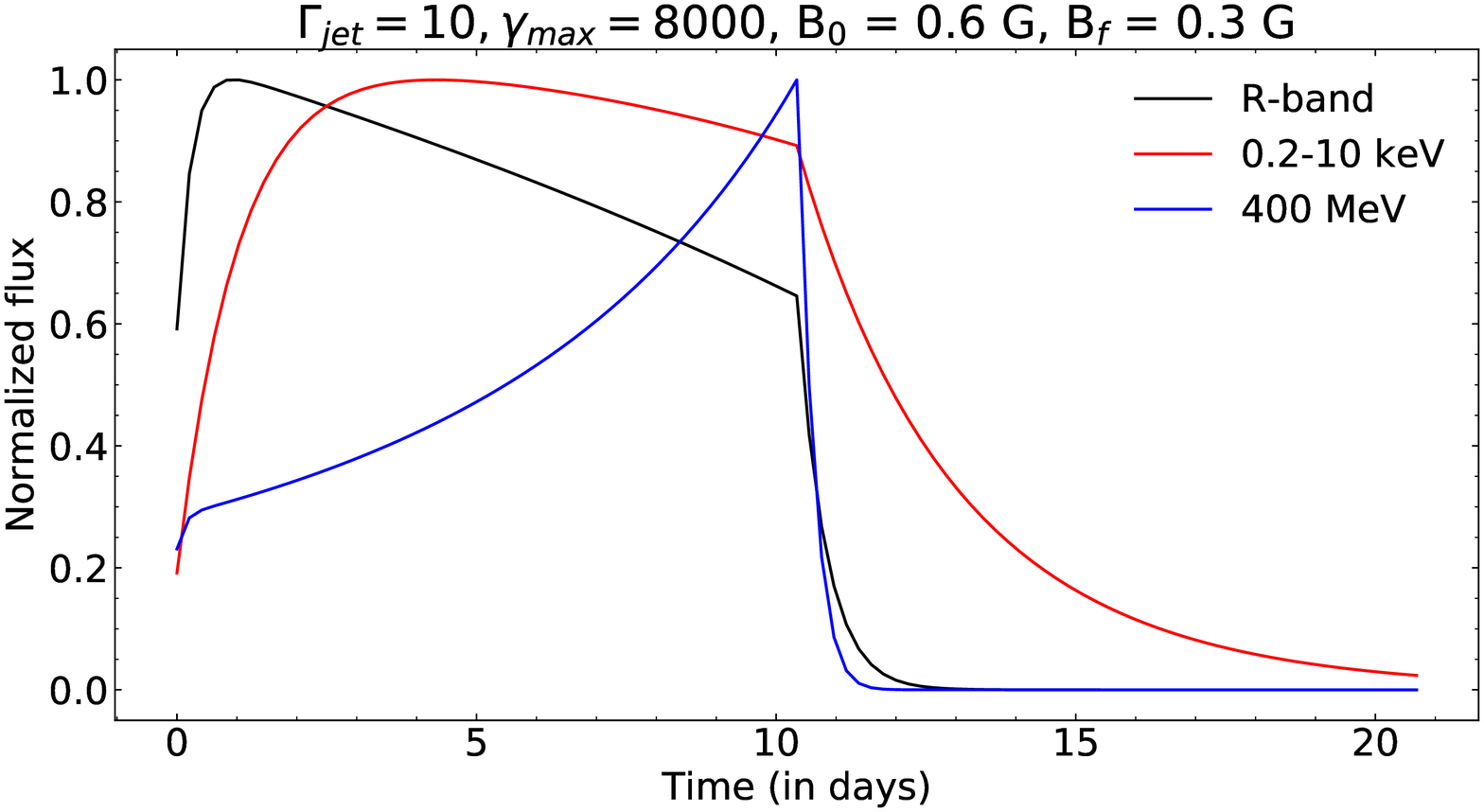}
	\caption{Light curves for an emission region of size $\sim 0.1$ pc which is very close to the torus ($\sim 9.5$ pc). The normalized flux reaches unity when the shock has energized all the cells.}
    \label{fig:near_torus}
\end{figure}

We note that the optical/$\gamma$-ray and X-ray/$\gamma$-ray delays depend on the size of the emission region as well as their relative distance to the BLR and torus. We restrict the size of the emission region so that the general shape of the light curves in our model resembles the observed light curves. However, in order to investigate the effects of the relative distances of the model emission region from the BLR and the torus, we run our model for $L_D = 10^{45}$ erg s$^{-1}$, $\Gamma_{jet} = 10$, $\gamma_{max} = 8000$, $B_0 = 0.6$ G and $B_f = 0.3$ G for an emission zone $\sim 9.5$ pc from the central black hole. We choose a high value of $\gamma_{max}$ so that the torus photons may be up-scattered to produce $400$ MeV emission. We show our results in Figure \ref{fig:near_torus}. We find that the $\gamma$-ray flux has a maximum at a much later time compared to the optical light curve. This is due to the fact that at such a large distance relative to the BLR, the BLR photon energy density in the jet frame is virtually non-existent due to the de-beaming effect. As a result, the total cooling rate of the electrons significantly drops. Hence, the electrons do not lose much energy and the number of electrons that can produce $\gamma$-rays continues to increase as the shock energizes more cells, resulting in a continuous increase in the $\gamma$-ray emission. After the shock leaves the emission region, the emission decays. At such large distances with respect to the BLR, we find that the optical/$\gamma$-ray time lag depends on the shock crossing time scale.

One of the features of our model is that it consistently predicts that $\gamma$-ray variations lag behind optical variations. But from the observational results, we find that it is possible for the former to lead the latter as well. The lead of the optical variability relative to the $\gamma$-rays depends on how much the magnetic field varies with distance compared to the external radiation density. In our model, the external radiation density depends weakly on distance for $r < r_{BLR}$, whereas the magnetic field has a much stronger dependence ($B \sim r^{-1}$). Therefore, the synchrotron radiation reaches its peak earlier compared to the $\gamma$-rays because the synchrotron radiation from the zones located downstream of the emission region (with smaller value of $B$) does not contribute significantly to the total optical emission. As a result, the dependence of the magnetic field on distance must be weaker than $r^{-1}$ in order for the theoretical optical variations to lag behind those at $\gamma$-ray energies. Another possibility is that the size of the emission region is much larger than the distance from the central engine to the BLR for the case of HSP blazars. Due to the de-beaming effect, the external energy density of the BLR photons in the jet frame, $u'_{BLR}$, decreases rapidly with distance inside the emission region. This is due to the $1+(r/r_{BLR})^{\beta_{BLR}}$ term in the denominator. As a result, the $\gamma$-ray emission should reach its maximum and start to decay faster than the optical emission. However, as $r_{BLR}$ for FSRQs is typically few times $0.1$ pc \citep{ghi11}, an emission region that is much larger (e.g., $> 1$ pc) will result into time delays of $10$ days or more. Hence, this argument cannot be used to explain time delays of a few days. 

When comparing the correlation results of observed light curves with our model, we work with an optical/$\gamma$-ray delay of $-1$ $\pm$ $2$ days, which does not include the case of the blazar PKS 0235+164, as it exhibits a positive delay of $16$ days. In our model, such a large delay does not seem to occur for the parameters that we explore unless the emission region is much larger than we have assumed based on flare shapes and timescales. This is because of the fact that to upscatter photons of energy $10$ eV ($\sim 10^{15}$ Hz) from the BLR to 400 MeV ($\sim 10^{23}$ Hz) requires electrons of Lorentz factor $\sim 10^3$, as the observed energy of radiation up-scattered by the electrons in the jet is $D^2 \gamma^2 h\nu_{ext}$ and the Doppler factor for our analysis is $\sim 10$. Optical emission is also produced by electrons having similar energies with approximately contemporaneous variations. Therefore, it is usually not expected that the optical/$\gamma$-ray variability will have a time delay of more than $10$ days. One of the possibilities is the so-called ``orphan'' flares in which a large outburst of optical emission is observed but no such flare at the $\gamma$-ray energies is detected or \textit{vice versa}. The latter can happen, e.g., if the jet is Compton-dominated as opposed to being magnetically dominated. In that case, there will be no optical flare relative to which the delay of the $\gamma$-ray variability may be measured. Therefore, these flares may appear to be correlated with the next or the previous optical flare, which may be more than $10$ days later. Alternatively, the emission region itself could be much larger than the one we have assumed in our simulation.

Finally, we note that the X-ray/$\gamma$-ray variability of most HSP blazars (which contain only HBL objects) does not show any correlation. This can be explained if $\gamma$-ray emission of these objects is dominated by the SSC and not the EC process. For BL Lac objects, the value of $L_{BLR}$ is significantly lower compared to the FSRQs and hence $u_{BLR}$ is lower accordingly. As a result, the SSC contribution to the total $\gamma$-ray emission is higher compared to that due to EC if $\gamma_{max}$ is high enough such that the former can generate $\gamma$-rays. For SSC-dominated emission, we do not expect much correlation as $\gamma$-rays of a given energy can be produced by up-scattering of seed photons by electrons of different energies. Alternatively, the lack of correlation may indicate other physical factors, such as a significant contribution of radiation from the host galaxy or generation of $\gamma$-rays due to hadronic processes.

\section{Summary}		\label{summary}
In this work we use the light curves of a sample of 26 unique blazars from \textit{Fermi-LAT}, \textit{Swift-XRT} and Yale-SMARTS blazar monitoring program to investigate the temporal correlation among their $\gamma$-ray, X-ray and optical variation. We find that:

\begin{enumerate}
	\item There is significant correlation among optical/$\gamma$-ray and X-ray/$\gamma$-ray light curves of LSP blazars in our sample. The X-ray/$\gamma$-ray correlation of HSP blazars on the other hand is weaker compared to that in the LSP blazars.  
	\item The average time delay in the optical/$\gamma$-ray DCCF of LSP blazars is $-1$ $\pm$ $2$ days. The X-ray/$\gamma$-ray average delay is found to be $-1$ $\pm$ $3$ days. For HSP blazars, the X-ray/$\gamma$-ray delay is found to be $2 \pm 2$ days. These values imply that no significant delay is present between the wavebands.
	\item In our theoretical model of nonthermal emission in the jet, we find that for LSP blazars, large changes in the magnetic field across the emission region ($> 0.5$ G) and very small values of magnetic fields (e.g., $0.2$ G) are consistently ruled out by our model. The observed nature of variability of HSP blazars, on the other hand, may be explained if the size of the emission region is much larger than the distance between the BLR and the central engine.
	\item Our model consistently predicts optical variability to lag behind that of the $\gamma$-ray emission while from observations we find that the opposite is also possible. We propose that this can be explained if the dependence of the magnetic field with respect to distance falls off at a rate slower than $r^{-1}$. We note that this can also happen in other scenario such as the presence of significant turbulence in the emission region. 
\end{enumerate}

\section*{Acknowledgements}

This study has made use of up-to-date SMARTS optical/near-infrared light curves\footnote{www.astro.yale.edu/smarts/glast/home.php} and near real-time results from Swift observations of Fermi-LAT ``sources of interest'' and flaring sources\footnote{http://www.swift.psu.edu/monitoring/}. $\gamma$-ray data are from the Fermi Science Support Center website\footnote{https://fermi.gsfc.nasa.gov/ssc/data/access/}. The authors acknowledge R. Corbet for his help regarding the use of the publicly available software like\_lc.pl, which has been used to carry out the LAT data analyses required for this work. AM receives the DST-INSPIRE fellowship. RC thanks Presidency University for support under the Faculty Research and Professional Development (FRPDF) Grant. RC acknowledges support from the UGC Start-Up Research Grant. RC thanks IUCAA for their hospitality and usage of their facilities during his stay at various times as part of the university associateship program. The authors thank the anonymous referee whose valuable comments led to improvement of this manuscript.



\bsp	
\label{lastpage}

\begin{thebibliography}{99}

\bibitem[Acero et al.(2005)]{3FGL} Acero, F., Ackermann, M., Ajello, M., et al. 2015, \apjs, 218, 23 

\bibitem[Arbeiter, Pohl \& Schlickeiser(2005)]{arb05} Arbeiter C., Pohl M., Schlickeiser R., 2005, \apj, 627, 62

\bibitem[B{\l}a\.{z}ejowski et al.(2000)]{bla00}B{\l}a\.{z}ejowski M., Sikora M., Moderski R., Madejski G.M., 2000, \apj, 545, 107

\bibitem[Bonning et al.(2012)]{bon12} Bonning E. W. et al., 2012, \apj, 756, 13

\bibitem[B{\"o}ttcher et al.(2010)]{bot10} B{\"o}ttcher M. et al., 2010, \apj, 725, 2344

\bibitem[Bregman et al.(1981)]{bre81}Bregman J. N., Lebofsky M.J., Aller M.F., Rieke G.H., Aller H.D., Hodge P.E., Glassgold A.E., Huggins P.J., 1981, \nat, 293, 714

\bibitem[Chatterjee et al.(2008)]{cha08} Chatterjee R. et al., 2008, \apj, 689, 79

\bibitem[Chiang \& B\"{o}ttcher(2002)]{chi02} Chiang J., B\"{o}ttcher M., 2002, \apj, 564, 92

\bibitem[Coppi \& Aharonian(1999)]{cop99} Coppi P. S., Aharonian F. A., 1999, \apj, 521, L33 

\bibitem[Dermer et al.(2009)]{der09} Dermer C.D., Finke J.D., Krug H., B{\"o}ttcher M., 2009, \apj, 692, 32

\bibitem[Edelson \& Krolik(1988)]{ede88} Edelson R. A., Krolik J. H., 1988, \apj, 333, 646

\bibitem[Fuhrmann et al.(2014)]{fuh14} Fuhrmann L. et al., 2014, \mnras, 441, 1899

\bibitem[Ghisellini et al.(2011)]{ghi11} Ghisellini G., Tavecchio F., Foschini L., Ghirlanda G., 2011, \mnras, 414, 2674

\bibitem[Hayashida et al.(2012)]{hay12} Hayashida M. et al., 2012, \apj, 754, 114

\bibitem[Impey \& Neugenbauer(1988)]{imp88} Impey C., Neugenbauer G., 1988, \aj, 95, 307

\bibitem[Jorstad et al.(2010)]{jor10} Jorstad S. G. et al., 2010, \apj, 715, 362

\bibitem[Maraschi, Ghisellini \& Celotti(1992)]{mar92} Maraschi L., Ghisellini G., Celotti A., 1992, \apj, 397, L5

\bibitem[Marscher(1998)]{mar98} Marscher A. P., 1998, in Zensus J.A., Wrobel J.M., Taylor G.M., eds, ASP Conf. Ser. Vol. 144, Radio Emission from Galactic and Extragalactic Compact Sources. Astron. Soc. Pacific, San Francisco, p. 25

\bibitem[Moderski, Sikora \& B{\l}a\.{z}ejowski(2003)]{mod03} Moderski R., Sikora M., B{\l}a\.{z}ejowski M., 2003, \aap, 406, 855

\bibitem[Sikora, Begelman \& Rees(1994)]{sik94} Sikora M., Begelman M.C., Rees M.J., 1994, \apj, 421, 153

\bibitem[Stroh \& Falcone(2013)]{str13} Stroh M. C.,  Falcone A. D., 2013, \apjs, 207, 28

\bibitem[Urry \& Mushotzky(1982)]{urr82} Urry C. M., Mushotzky R. F., 1982, \apj, 253, 38 

\bibitem[Urry \& Padovani(1995)]{urr95} Urry C. M., Padovani P., 1995, \pasp, 107, 803

\end{thebibliography}
\end{document}